# Quantum SASE FEL with laser wiggler

Rodolfo Bonifacio*

INFN – Sezione di Milano and Laboratorio Nazionale di Frascati (LNF), Italy

and

Departamento de Fisica, Universidade Federal Alagoas, Maceiò, 57072-970, Brazil

Abstract

In this letter we specify the physical parameters necessary to operate a SASE FEL in the quantum regime with a laser wiggler. We also show that this is more feasible in the quantum regime than in the classical one. Specific examples are given.

*Corresponding author. Fax: +55 82 2931844

E-mail address: rodolfo.bonifacio@tin.it

Departamento de Física, Universidade Federal de Alagoas

Campus A. C. Simões, BR 104 km 14, Tabuleiro dos Martins , 57072-970 Maceió-AL, Brazil

It has been previously recognized that quantum effects in SASE FEL are determined by the quantum FEL parameter $\bar{\rho}$ [1] defined as the usual FEL parameter times the ratio between the electron energy and the photon energy. Quantum effects become relevant when $\bar{\rho}$ < 1. However the calculations of Ref. [1] are confined to the linear regime. In Ref. [2] we have extended the theory of Quantum SASE FEL to the non linear regime and we have shown the phenomenon of quantum purification of SASE spectrum, i.e., the broad superposition of chaotic series of random spikes predicted by the classical theory shrinks to a very narrow spectrum of the emitted radiation when $\bar{\rho}$ <1. The question is: what is the experimental set up and the experimental parameters necessary to observe Quantum SASE in the short wavelength region (where quantum effects are expected and more relevant)?

The first possibility is the usual configuration of GeV accelerators and very long magnetic undulators as in the SLAC LCLS and DESY TESLA-FEL projects. However the quantum regime would require even longer magnetic undulator, because to reach the quantum regime one needs a very low value of the FEL parameter, which determines the gain per wiggler period.

Alternatively one can propose the use of a typical Compton back scattered configuration: a low energy electron beam counter-propagating with respect to an electromagnetic wiggler (wave) provided by a high power laser. In this note we propose the Compton configuration giving the scaling laws and the expressions of all relevant physical quantities as a function of $\bar{\rho}$, both in the quantum and the classical case. In particular, we show that this experimental set up appears reasonable in the quantum SASE regime $\bar{\rho}$ << 1, whereas it is much more problematic in the classical regime, due to fact that the scaling laws as a function of $\bar{\rho}$ are much different in the two cases.

Let us consider a laser wiggler with radiation propagating in the z direction opposite to an electron beam with the following specifications: $W_0$ is the minimum diameter of the laser beam in the focus, $\sigma_0$ is the minimum radius of the electron beam, $Z_0$ is the distance in which the radiation beam diverges (Rayleigh range), $\beta^*$ is the analogous length for the electron beam, $\gamma$ is the Lorentz factor, $\varepsilon_n$ is the normalized beam emittance and $\lambda_r$ is the FEL radiation wavelength length. The following relations are well known [3]:

(1) $$Z_0 = \frac{\pi W_0^2}{\lambda}$$

(2) $$\sigma_0 = \sqrt{\frac{\varepsilon_n \beta^*}{\gamma}}$$

In order to ensure a good overlap and matching between radiation and electron beam we must impose that

(3) $$\varepsilon_1 = \frac{W_0}{4\sigma_0} \geq 1$$

(4) $$\varepsilon_2 = \frac{\beta^*}{Z_0} \geq 1$$

Condition (3) is rather conservative [4], but ensures that the laser intensity is almost constant over the electron beam transverse profile. If the wiggler parameter is small condition (3) can be relaxed. The FEL resonance condition for an electromagnetic wiggler reads:

(5) $$\gamma = \frac{1}{2}\sqrt{\frac{\lambda}{\lambda_r}(1+a_w^2)}$$

We remark that for a magnetic wiggler the factor $1/2$ should be replaced by $1/\sqrt{2}$.

Imposing the consistency of (1) - (4) and using Eq. (5) we obtain:

(6) $$\lambda_r(A) = \lambda^3 \frac{(1+a_w^2)}{(32\pi\eta)^2} = \frac{\lambda^3(\mu m)}{\eta^2}(1+a_w^2)$$

where

(7) $\quad \eta = \varepsilon_1^2 \varepsilon_2 \varepsilon_n (mm.mrad)$.

We remark that Eq. (6) (formally independent on the electron energy) gives a direct relation between the radiation wavelength and the wiggler wavelength in terms of two geometrical parameters and $\varepsilon_n$ (via the $\eta$ factor) and the wiggler parameter. The previous relations can be derived using the following chain of equations:

$$Z_0 = \frac{\pi W_0^2}{\lambda} = \frac{16\pi\varepsilon_1^2 \sigma_0^2}{\lambda} = \frac{16\pi\varepsilon_1^2 \varepsilon_n B^*}{\lambda\gamma} = \frac{32\pi\varepsilon_1^2 \varepsilon_2 \varepsilon_n}{\lambda^{3/2}} \left(\frac{\lambda_r}{1+a_w^2}\right)^{1/2} Z_0.$$

Eliminating $Z_0$ from the first and the last equation, we obtain Eqs. (6) and (7). As an example if we take, $\varepsilon_1=\varepsilon_2=\varepsilon_n=1$, $\lambda_r=1$A, and $a_w \ll 1$, we obtain $\lambda=1$ μm.

Using Eq. (5), Eq. (6) can be written as

(8) $\quad \gamma = \frac{16\pi\eta}{\lambda} = 50\frac{\eta}{\lambda(\mu m)}$

Equation (8) fix the resonant energy only in terms of the parameter $\eta$ and of the laser wiggler wavelength. For example for $\eta =1$ and $\lambda = 1$ μ one has $\gamma = 50$.

Let us remember the definition of the quantum FEL parameter [1,2]:

(9) $\quad \overline{\rho} = \rho_F \frac{mc\gamma}{\hbar k_r} = \gamma\rho_F \frac{\lambda_r}{\lambda_C}$.

where $\lambda_c = \frac{h}{mc} \simeq 0.024$ A is the Compton wavelength. Inverting Eq. (9) and using Eq. (6) one has

(10) $\quad \rho_F = 5\cdot 10^{-4} \frac{\eta}{\lambda(\mu m)^2} \frac{\overline{\rho}}{1+a_w^2}$.

Using Eq. (10), we can write the power gain length as:

(11) $\quad L_g[\mu m] = \frac{\lambda[\mu m]}{8\pi\rho_F \sqrt{\overline{\rho}}}\left(\sqrt{1+\overline{\rho}}\right) = 83\frac{\lambda^3[\mu m]}{\eta\overline{\rho}^{3/2}}\left(\sqrt{1+\overline{\rho}}\right)\left(1+a_w^2\right)$.

Equation (11) is not an exact equation, but is an interpolation formula which gives the correct behaviour in the quantum regime $\bar{\rho} \ll 1$ [2] and the classical expression [5] in the opposite limit. This equation can be rigorously justified in the asymptotic cases $\bar{\rho}$ very large or very small.

We must also impose that the electron beam characteristic length β* is larger than the gain length, i.e.,

$$\text{(14)} \quad \frac{\beta^*}{L_g} = \varepsilon_3 \geq 1$$

Hence, using Eqs. (1) - (4) and (15), one has:

$$\text{(15)} \quad \sigma_0^2[\mu m] = \frac{\varepsilon_n \beta^*}{\gamma} = \frac{\varepsilon_n \varepsilon_3 L_g}{\gamma} = 1.66 \varepsilon_n \varepsilon_3 \frac{\lambda^4[\mu m]}{\eta^2 \bar{\rho}^{3/2}} \left(\sqrt{1+\bar{\rho}}\right)\left(1+a_w^2\right).$$

Furthermore, using Eq. (3) and (15), it can be easily shown that:

$$\text{(16)} \quad a_w \simeq B_w(T)\lambda(cm) \simeq K \frac{\lambda}{W_0}\sqrt{P[TW]} \simeq \frac{0.2K}{\lambda[\mu m]} \sqrt{\frac{\eta \varepsilon_2}{\varepsilon_3}} \left(\frac{P[TW]\bar{\rho}^{3/2}}{(1+a_w^2)\left(\sqrt{1+\bar{\rho}}\right)}\right)^{1/2}$$

where P is the laser power, $B_w$ is the r.m.s. value of the laser magnetic field and $K \approx 5$, if the e-beam has a gaussian transverse profile [4], or K = 7 for a flat transverse profile with the same total power and beam waist.

Equation (16) is a self consistent equation for $a_w$, i.e.,

$$a_w^2 = \frac{a_0^2}{1+a_w^2}$$

where:

$$\text{(17)} \quad a_0^2 \simeq 4 \cdot 10^{-2} K^2 P \frac{\varepsilon_2}{\varepsilon_3} \frac{\eta}{\lambda^2} \left(\frac{\bar{\rho}^{3/2}}{\sqrt{1+\bar{\rho}}}\right)$$

is the wiggler parameter when $a_w \ll 1$.

Solving the previous equation we obtain easily

(18) $$a_w = \frac{a_0}{\sqrt{F(a_0)}}$$

where

(19) $$F(a_0) = \frac{1+\sqrt{1+4a_0^2}}{2} = 1 + a_w^2 > 1 .$$

Note that in the limit $4a_0^2 \ll 1$, $a_w \simeq a_0$, whereas in the opposite limit $a_w \simeq \sqrt{a_0}$.

Furthermore, the relative energy spread is subjected to the limitation

(20) $$\frac{\Delta\gamma}{\gamma} \leq \frac{4\rho_F\sqrt{\bar{\rho}}}{\left(\sqrt{1+16\bar{\rho}}\right)} = \frac{2\cdot 10^{-3}}{F(a_0)} \frac{\eta}{\lambda^2[\mu m]} \frac{\bar{\rho}^{3/2}}{\left(\sqrt{1+16\bar{\rho}}\right)}$$

where Eq. (10) and (19) has been used.

Equation (20) is NOT an exact expression but it gives an interpolation formula which gives the correct expression in the quantum limit $\bar{\rho} \ll 1$ and the classical expression [5] in the opposite limit. This equation can be rigorously justified in the asymptotic cases $\bar{\rho}$ very large or very small.

Substituting $\gamma\rho_F = 0.136 J^{1/3} B_w^{2/3} \lambda^{4/3} (S.U.)$ [5] in Eq. (9), we get

$$\bar{\rho}^3 = 2\cdot 10^{-2} J(A/\mu m^2) a_w^2 \lambda^2(\mu m) \lambda_r^3(A) .$$

Using Eq. (15) and the fact that $J=I/\pi\sigma_0^2$, the previous equation becomes

$$\bar{\rho}^3 = 1.44\cdot 10^{-4} \cdot K^2 \frac{\bar{\rho}^3}{\left(\sqrt{1+\bar{\rho}}\right)^2} PI\lambda^5 \left(\frac{\varepsilon_2}{\varepsilon_n \varepsilon_3^2 \eta^3}\right) F(a_0) .$$

Hence, we obtain:

(21) $$I = \frac{7\cdot 10^3}{K^2} \frac{1}{P\lambda^5} \left(\frac{\varepsilon_n \varepsilon_3^2 \eta^3}{\varepsilon_2 F(a_0)}\right) \left(\sqrt{1+\bar{\rho}}\right)^2 .$$

We remark that in the quantum limit $\bar{\rho} \ll 1$ the current is independent on $\bar{\rho}$, whereas in the opposite limit it increases as $\bar{\rho}$.

The minimum laser time duration required is given by:

$$(22) \quad \tau(p\sec) = \frac{L_g}{c} = 3.3 \cdot 10^{-3} \varepsilon_3 L_g [\mu m] = 0.3 \varepsilon_3 \left(\sqrt{1+\bar{\rho}}\right) \frac{\lambda^3 F(a_0)}{\eta \bar{\rho}^{3/2}}.$$

where

$$(23) \quad L_g = 83 \frac{\lambda^3 \left(\sqrt{1+\bar{\rho}}\right)}{\eta \bar{\rho}^{3/2}} F(a_0).$$

Furthermore, we have

$$(24) \quad \lambda_r = \frac{\lambda^3}{\eta^2} F(a_0),$$

$$(25) \quad E(MeV) = 25 \frac{\eta}{\lambda},$$

$$(26) \quad \frac{\Delta\gamma}{\gamma} \leq \frac{2 \cdot 10^{-3}}{\left(\sqrt{1+16\bar{\rho}}\right)} \frac{\eta}{\lambda^2} \frac{\bar{\rho}^{3/2}}{F(a_0)}.$$

$$(27) \quad \sigma_0^2[\mu m] = \frac{\varepsilon_n \beta^*}{\gamma} = \frac{\varepsilon_n \varepsilon_3 L_g}{\gamma} = 1.66 \varepsilon_n \varepsilon_3 \frac{\lambda^4 \left(\sqrt{1+\bar{\rho}}\right)}{\eta^2 \bar{\rho}^{3/2}} F(a_0)$$

The units are: $\lambda$ in $\mu$m, P(TW), $\lambda_r$ in Angstrom. The other characteristic lengths $\beta^*$, $Z_0$, and $W_0$ can be obtained by Eqs. (1-4) and (27).

Two explicit examples for a laser wiggler at 1 $\mu$m and 10 $\mu$m are given in Table 1, where all the other physical parameters are expressed as a function of the laser power, P. The values in parenthesis are for P=1TW.

In conclusion, Eqs. (17)-(27) and relations (1)-(4) give the relevant quantities for the design of a SASE FEL in terms of the quantum FEL parameter $\bar{\rho}$, the pump laser wavelength $\lambda$, the unperturbed wiggler parameter $a_0$, and of adimensional parameters $\eta$ (Eq. (7)). The quantum SASE regime is obtained when $\bar{\rho}$< 1 whereas in the opposite limit the classical regime with a laser wiggler is recovered. We remark again that Eq. (21) shows that in the classical limit, where $\bar{\rho}$>>1, the required current increases as $\bar{\rho}$ and is much larger than in the quantum regime, where it is independent of $\bar{\rho}$. This fact makes more problematic the use of an electromagnetic wiggler in the classical regime than in the quantum regime. If the quantum regime with the laser wiggler is experimentally feasible, a short wavelength **coherent** FEL can be a table top object. In such a case the technological problem would go from high energy accelerators plus long magnetic wigglers to the widely used high power laser technology.

This work has been completely supported by INFN, Sezione di Milano and Frascati.

The author is grateful to Dr. Massimo Ferrario, Dr. Nicola Piovella and Dr. Luca Serafini for helpful discussion and suggestions and Dr. Lucia de Salvo for continuous assistance. The author would like also to thank Prof. Sergio Bertolucci and Prof. Luigi Palumbo for their support and interest.

| λ (μm) | 1 | 10 |
|---|---|---|
| $\varepsilon_1$ | 1 | $\sqrt{10}$ |
| $\varepsilon_2$ | 1 | 10 |
| $\lambda_r (Angstrom)$ | $F(P) \simeq (1)$ | $0.1 F(P) = (0.15)$ |
| $a_0$ | $0.28\sqrt{P}$ | $0.9\sqrt{P}$ |
| $F(P)$ | $\left(1+\sqrt{1+0.31P}\right)/2 \simeq (1)$ | $\left(1+\sqrt{1+3.5P}\right)/2 \simeq (1.5)$ |
| $a_w$ | $0.3\sqrt{P/F(P)} \simeq (0.3)$ | $0.9\sqrt{P/F(P)} = (0.7)$ |
| $I(Amp)$ | (312) | (218) |
| $E(MeV)$ | 25 | 250 |
| $\Delta\gamma/\gamma \leq$ | $3.4\cdot10^{-4}/F(P) \simeq (3.4\cdot10^{-4})$ | $3.7\cdot10^{-4}/F(P) \simeq (2.4\cdot10^{-4})$ |
| $\tau(p\sec)$ | 4 | 56 |
| $d(\mu m) = 2\sigma_0$ | $9.4 F(P) \simeq (9.4)$ | $7.2 F(P) = (11)$ |
| $W_0(\mu m)$ | $18 F(P) = (18)$ | $0.5 F(P) = (70)$ |
| $Z_0(\mu m) = \beta*/\varepsilon_2$ | $10^3 \cdot F^2(P) = (10^3)$ | $7\cdot10^2 \cdot F^2(P) = (1.6\cdot10^3)$ |

Table 1. Examples of the various parameters for $\varepsilon_3$=1 e $\varepsilon_n$=1, K=5, $\bar{\rho} = 0.2$. Here the unique free parameter is the laser pump power, P (TW). The number in parenthesis are for P = 1TW.